\documentclass[structabstract]{aa}
\usepackage{amssymb,amsmath}
\usepackage{graphicx}
\usepackage{txfonts}
\usepackage{natbib}  
\usepackage{color} 
\bibpunct{(}{)}{;}{a}{}{,} 
\newcommand{\CII}{\ion{C}{II}}
\newcommand{\OI}{\ion{O}{I}}
\newcommand{\CIIfs}{[\ion{C}{II}]}
\newcommand{\OIfs}{[\ion{O}{I}]}  
\begin{document}
\title{Nature of the gas and dust around 51 Oph\thanks{{\it Herschel}
    is an ESA space observatory with science instruments provided by
    European-led Principal Investigator consortia and with important
    participation from NASA.}}  \subtitle{Modelling continuum and
  Herschel line observations} \author{W. F. Thi\inst{1},
  F. M\'{e}nard\inst{2,1}, G. Meeus\inst{3}, A. Carmona\inst{1},
  P. Riviere-Marichalar\inst{4,3}, J.-C. Augereau\inst{1},
  I. Kamp\inst{5}, P. Woitke\inst{6}, C. Pinte\inst{1},
  I. Mendigut\'{i}a\inst{7}, C. Eiroa\inst{3}, B.  Montesinos\inst{3},
  S. Britain\inst{7}, W. Dent\inst{8}}

\titlerunning{The disc around \object{51 Oph}} 
\institute{UJF-Grenoble
  1 / CNRS-INSU, Institut de Plan\'{e}tologie et d'Astrophysique
  (IPAG) UMR 5274, Grenoble, F-38041, France
  \email{Wing-Fai.Thi@obs.ujf-grenoble.fr} \and UMI -- LFCA, CNRS / INSU
  France, and Dept. de Astronomia y Obs. Astronomico Nacional,
  Universidad de Chile, Casilla 36-D, Correo Central, Santiago, Chile
  (UMI 3386)\and Dep. de F\'isica Te\'orica, Fac. de Ciencias, UAM
  Campus Cantoblanco, 28049 Madrid, Spain \and Centro de Astrobiologıa
  – Depto. Astrofısica (CSIC–INTA), ESAC Campus, P.O. Box 78, 28691
  Villanueva de la Canada, Spain \and Kapteyn Astronomical Institute,
  P.O. Box 800, 9700 AV Groningen, The Netherlands \and SUPA, School
  of Physics \& Astronomy, University of St. Andrews, North Haugh,
  St. Andrews KY16 9SS, UK \and Department of Physics and Astronomy,
  Clemson University, USA \and ALMA, Avda Apoquindo 3846, Piso 19,
  Ediﬁcio Alsacia, Las Condes, Santiago, Chile}
\authorrunning{W.-F. Thi}

   \date{Received 2012; accepted 16 April 2013}

 
  \abstract
  {Circumstellar disc evolution is paramount for the understanding of
    planet formation. The gas in protoplanetary discs large program
    (GASPS) aims at determining the circumstellar gas and solid mass
    around $\sim$~250 pre-main-sequence Herbig~Ae and T~Tauri stars.}
  {We aim to understand the origin and nature of the circumstellar
    matter orbiting \object{51 Oph}, a young ($<$1 Myr) luminous B9.5
    star.}
  {We obtained continuum and line observations with the PACS
    instrument on board the Herschel Space Observatory and continuum
    data at 1.2~mm with the IRAM 30m telescope. The spectral energy
    distribution and line fluxes were modelled using the physico-chemo
    radiative transfer code ProDiMo to constrain the gas and solid
    mass of the disc around \object{51~Oph}. The disc vertical
    hydrostatic structure was computed self-consistently together with
    the gas thermal balance.}
  {We detected a strong emission by atomic oxygen \OIfs\ at 63 microns
    using the Herschel Space Observatory. The \OIfs\ emission at 145
    microns, the \CIIfs\ emission at 158 microns, the high-$J$ CO
    emissions, and the warm water emissions were not
    detected. Continuum emission was detected at 1.2~mm. The continuum
    from the near- to the far-infrared and the \OIfs\ emission are
    well explained by the emission from a compact
    ($R_{\mathrm{out}}$=10--15~AU) hydrostatic disc model with a gas
    mass of 5$\times$10$^{-6}$ M$_\odot$, 100 times that of the solid
    mass. However, this model fails to match the continuum millimeter
    flux, which hints at a cold outer disc with a mass in solids of
    $\sim$~10$^{-6}$ M$_\odot$ or free-free emission from a
    photoevaporative disc wind. This outer disc can either be devoid
    of gas and/or is to cold to emit in the \OIfs\ line. A very flat
    extended disc model ($R_{\mathrm{out}}$=400~AU) with a fixed
    vertical structure and dust settling matches all photometric
    points and most of the \OIfs\ flux.}
    {The observations can be explained by an extended flat disc
    where dust grains have settled. However, a flat gas disc cannot be
    reproduced by hydrostatic disc models. The low mass of the
    \object{51~Oph} inner disc in gas and dust may be explained either
    by the fast dissipation of an initial massive disc or by a very
    small initial disc mass.}  


   \maketitle
%
\section{Introduction}\label{introduction}

The large variety of extrasolar planetary systems requires a
theoretical and observational understanding of protoplanetary discs
where the exoplanets are formed. Observing protoplanetary discs at
different stages of evolution makes it possible to test the various
models of disc dissipation, planet-formation, and
planet-migration. Discs are characterized by their gas and solid
contents and their geometry. The two main categories of discs are the
young massive and flaring ones
($M_{\mathrm{disc}}$=10$^{-2}$-10$^{-3}$ M$_\odot$ in gas and dust,
typical size of 100-500 AU), characterized by a strong IR excess in
their spectral energy distribution (SED); and the older debris discs,
where there is no gas left and where the small dust grains result from
the collisions of planetesimals. The dust discs in the latter category
are geometrically flat and the SED shows weak excess in the
far-IR. The passage from one to the other categories seems rapid
because there are very few of the so-called transition discs. The
definition of a transition disc is not commonly accepted yet, but
discs with a lack of near-IR excess because of an inner hole caused by
a planet but that have strong far-IR excess can be considered as
transitional discs.  The rapidly rotating southern B9.5IV-B9.5V star
\object{51 Oph} (a.k.a. \object{HR 6591}, \object{HD 158643},
\object{SAO 185470}, \object{IRAS 17283-2355}) located at $\sim$131 pc
is a rare example of a pre-main-sequence star that is surrounded by a
disc where most or all of the primordial gas has probably
dissipated. The SED of the \object{51 Oph} disc suggests a new
category of disc. It does show a near-IR excess, but the main property
of the SED is a drop in the continuum flux beyond 20 microns. No
observations existed beyond 60 microns. The absence of strong far-IR
continuum dust emission points to a small amount of dust grains and a
non-flaring disc.  The 10-microns feature is consistent with emission
by amorphous silicates and, there are almost no features that can be
attributed to polycyclic aromatic hydrocarbon (PAH) emissions
\citep{Keller2008ApJ...684..411K}. \citet{Stark2009ApJ...703.1188S}
proposed a two-optically thin disc model to explain the combination of
MIDI-VLT, Keck nulling, and {\it Spitzer-IRS} data. Their outer dust
disc extends to 1200~AU and is flat.
\begin{figure*}[!ht]
\centering
\resizebox{\hsize}{!}{\includegraphics[angle=90]{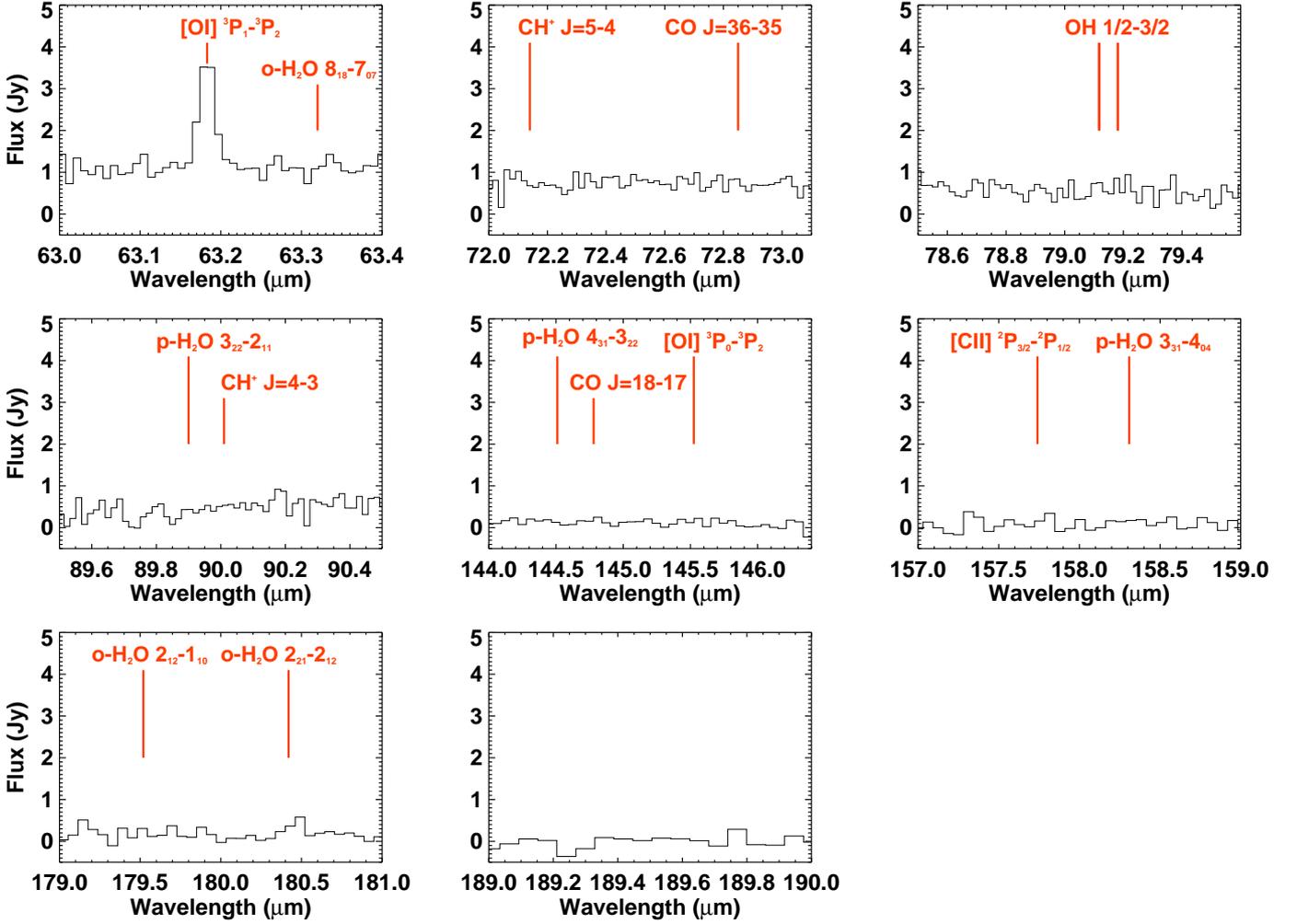}}
\caption{Herschel spectra towards \object{51 Oph}. The \OIfs\ line is spectrally unresolved. The line peak wavelengths are shown.} 
  \label{fig_herschel_spectra}          
\end{figure*}    
      
The evolutionary status of \object{51 Oph} is debated. The low dust
mass favours the idea that \object{51 Oph} is a young debris disc as
reported in \cite{Stark2009ApJ...703.1188S}. A debris disc would have
a lower gas-to-dust mass ratio than a young disc. One of the key
questions is therefore how much gas (atomic + molecular) is still
present in the \object{51 Oph} disc? The total atomic + molecular gas
mass of the \object{51 Oph} disc has never been determined. In
addition to the mass, the nature of the gas in the 51 Oph disc is also
disputed. Is the gas atomic or molecular? A predominately atomic disc
would favour the scenario where most of the gas comes from the
evaporation of infalling comets without primordial gas left, whereas
significant amount of molecular gas (H$_2$) as traced by rotational
emissions by CO and its isotopologues would indicate that the gas is a
leftover from planet-formation and/or disc dissipation. Absorption
studies in the ultraviolet with the FUSE telescope
\citep{Grady1993ApJ...402L..61G,Lecavelier1997A&A...321L..39L,Roberge2002ApJ...568..343R}
suggest infalling atomic gas and no molecular gas, similar to the case
of the \object{$\beta$ Pic} debris disc.  On the other hand, spectra
in the infrared show ro-vibrational emission lines from molecules such
as CO, CO$_2$, and H$_2$O
\citep[e.g.][]{vandenAncker2001A&A...369L..17V}. The presence of
molecules suggests that the gas may be dense and warm enough so that
efficient gas-phase neutral-neutral reactions can produce molecules
such as water. Neutral carbon (\ion{C}{I}) has been detected in
absorption with the {\it FUSE} telescope, and the \ion{C}{I}/dust
column density ratio is low, similar to the value found in the
line-of-sight of \object{$\beta$ Pic}. Hot carbon monoxide (CO
bandhead emission) has been observed by several groups and was
assigned to the emission from a hot gas in a disc seen close to
edge-on
\citep{Thi2005A&A...430L..61T,Berthoud2007ApJ...660..461B,Tatulli2008A&A...489.1151T},
but no rotational emission from cold ($T$=10-30 K) CO in the
sub-millimeter range is available.  In this paper, we present new
observations in the millimeter and far-IR obtained with the {\it IRAM
  30m} telescope and the {\it Herschel-PACS Space Observatory} in
Sec.~\ref{observations} \&~\ref{obs_results}. The \object{51~Oph} {\it
  Herschel data} is part of the HerbigAe and debris disc survey from
the {\it GASPS} large open time programme (P.I. Dent; Dent et al., in
press). The new observations (continuum and line) combined with
archival data are modelled to derive an estimate of the disc dust and
gas masses in Sec.~\ref{modelling}.  In the discussion, we consider
the nature and the different possible origins of the \object{51 Oph}
disc (Sec.~\ref{discussion}). 

\section{Observations and data reduction}\label{observations}

The source \object{51 Oph} was observed for continuum emission at 1.2
mm using the MAMBO2 bolometer array \citep{Kreysa1998SPIE.3357..319K}
on the {\it IRAM 30m} telescope at Pico Veleta, Spain. Observations
were conducted during the bolometer pool in 2008 (Nov. 13-18, 23).
Zenith opacity for our observations was typically
$\sim$0.2--0.3. Observations were first carried out to a target
1~$\sigma$ sensitivity of 1 mJy (20 minutes on-source) and then
repeated on two subsequent nights to reach 5 sigma detection. A total
of 60 minutes on source was achieved, in an ON-OFF pattern of 1 minute
on target followed by 1 minute on sky, with a throw of 32\arcsec. Flux
calibration was carried out using Mars, Saturn, or Jupiter (depending
on availability), and local pointing and secondary flux calibration
was carried out using \object{IRAS 16293-2422B}. The data were reduced
using the facility reduction software, MOPSIC
\footnote{http://www.iram.es/IRAMES/mainWiki/CookbookMopsic}.

We obtained photometric and line observations with the {\it PACS}
instrument \citep{Poglitsch2010A&A...518L...2P} on board the {\it
  Herschel Space Observatory} \citep{Pilbratt2010A&A...518L...1P}.  We
obtained PACS scan map observations at 70, 100, and 160 microns.  At
100 microns, \object{51 Oph} was observed at two angles (70 and 110
degrees) to improve the noise suppression, while at 70 microns, we
only observed one scan direction. Both the 70 and 100 microns scans
include the red band at 160 microns. The PACS photometric data were
reduced with the mini scanmap pipeline in HIPE version 8.1.0, (calTree
version 32), and we constructed mosaics of the two scan maps at 100
and 160 microns. The continuum flux was extracted using an aperture of
12" (sky between 30 and 35") at 70 microns, of 15" (sky between 33 and
38") at 100 microns, of 20" (sky between 40 and 50") at 160
microns. We obtained fluxes of 873$\pm$23 mJy at 70, 347$\pm$10 mJy at
100 and 90$\pm$6 mJy at 160 microns (see
Table~\ref{table_results}). The calibration errors haven been
included.  The spectroscopic data were reduced with the official
version 8.0.1 of the Herschel Interactive Processing Environment
(HIPE; \citealt{Ott2010ASPC..434..139O}), using standard tasks
provided in HIPE. These include bad pixel flagging; chop on/off
subtraction; spectral response function division; rebinning with
oversample = 2 and upsample =1, corresponding to the native resolution
of the instrument; spectral flatfielding and finally averaging of the
two nod positions. To conserve the best signal and to not
introduce additional noise, we only extracted the central spaxel and
corrected for the flux loss with an aperture correction provided by
the PACS instrument team.  We extracted the fluxes of the detected
lines using a Gaussian fit to the emission lines with a first-order
polynomial to the continuum, using the RMS on the continuum (excluding
the line) to derive a 1 $\sigma$ error on the line by integrating a
Gaussian with height equal to the continuum RMS and width of the
instrumental FWHM. In case of a non-detection, we give a 3 $\sigma$
upper limit, also calculated on the continuum RMS. The measured line
fluxes are listed in Table~\ref{table_results}.

Pipeline-processed FEROS \footnote{The Fiber-fed Extended Range
  Optical Spectrograph, Kaufer, A. et al. 1999, The Messenger 95, 8}
ESO 2.2m high-resolution (R$\sim$ 45000) optical spectra (4000-9000
\AA) were obtained from the Advanced Data Products ESO
archive\footnote{http://archive.eso.org/}.  Details of the FEROS data
reduction pipeline can be found in the ESO HARPS and FEROS pipeline
web-site
\footnote{http://www.eso.org/sci/facilities/lasilla/instruments/harps/tools/
  software.html/}.  The spectra are displayed at the observed
wavelength and no correction was performed for the radial velocity or
motion of the earth around the sun. Finally, we complemented the new
data with archival photometry.
\begin{figure}[!ht]
\centering
\resizebox{\hsize}{!}{\includegraphics[]{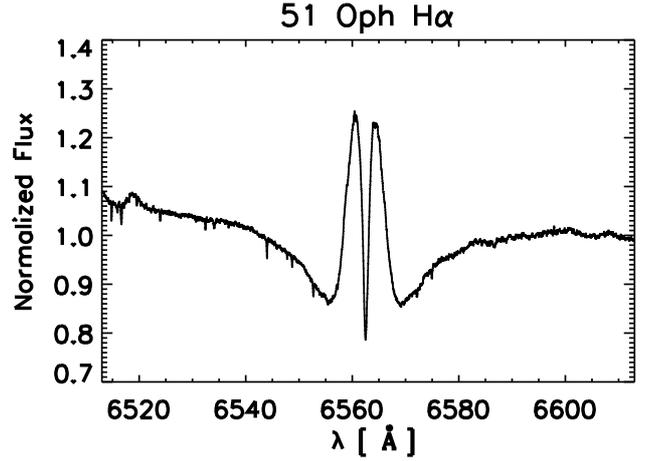}}
\caption{Normalized optical spectrum of \object{51 Oph} around the
  H$\alpha$ emission obtained by FEROS.}
  \label{fig_Halpha_spectrum}          
\end{figure}    
\begin{figure}[!ht]
\centering
\resizebox{\hsize}{!}{\includegraphics[]{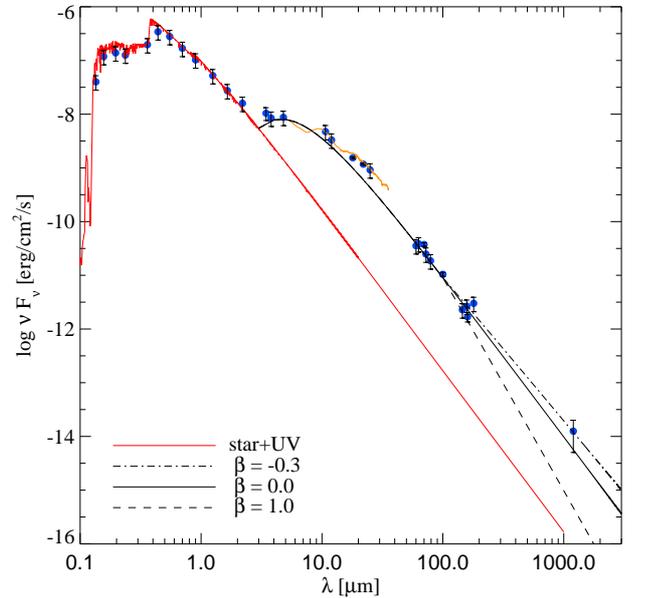}}
\caption{Modified blackbody fits to the \object{51 Oph} SED. The error bars on the photometric points are 3~$\sigma$. The orange line is the {\it Spitzer-IRS} spectrum taken from the NASA archive. The best fit with $\beta$=-0.3 is unphysical.}
  \label{fig_sed_modified_bb}          
\end{figure}    

\section{Observational results}\label{obs_results}

\begin{figure*}[!ht]
\centering
 \includegraphics[scale=0.5]{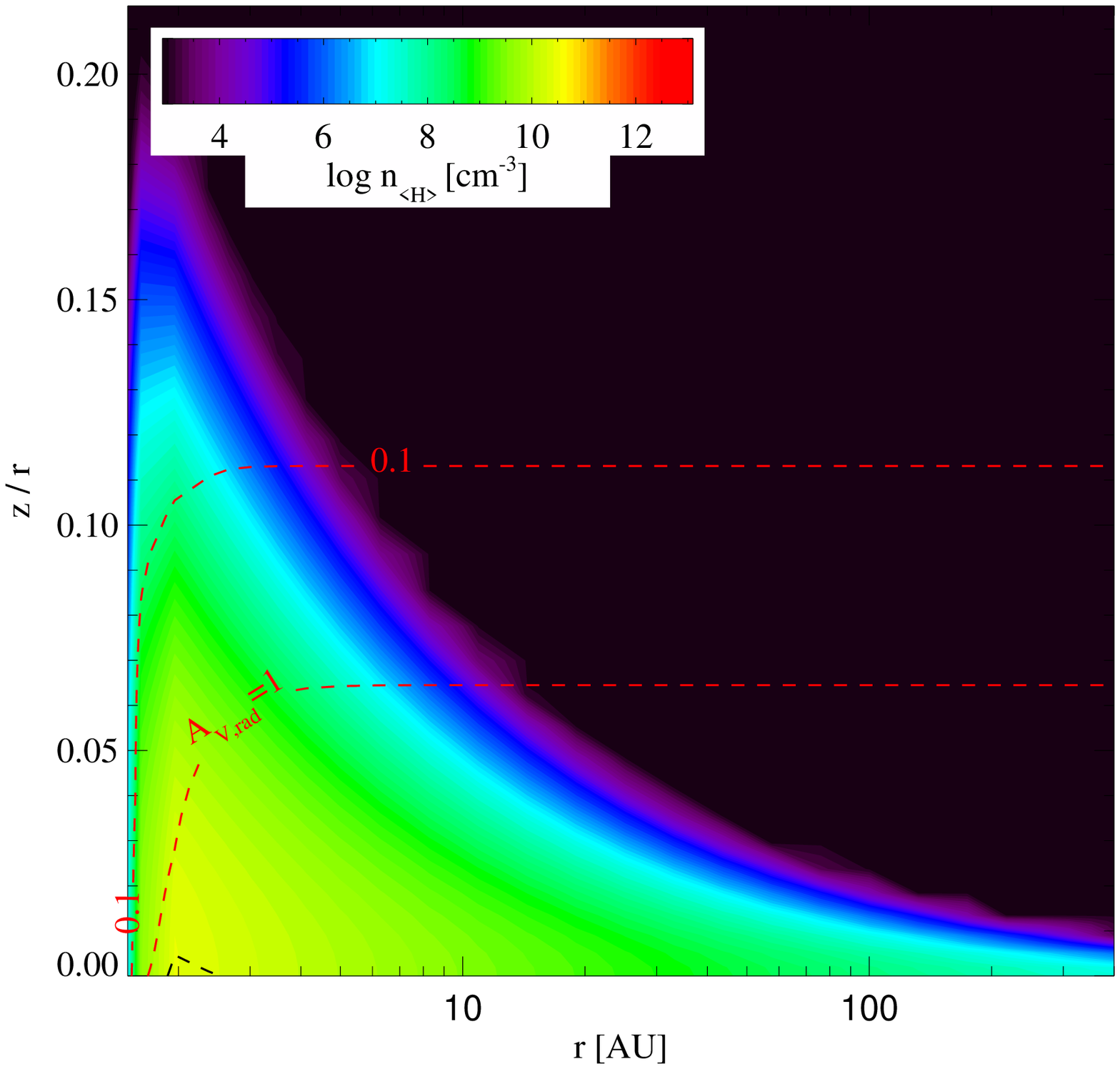}
 \includegraphics[scale=0.5]{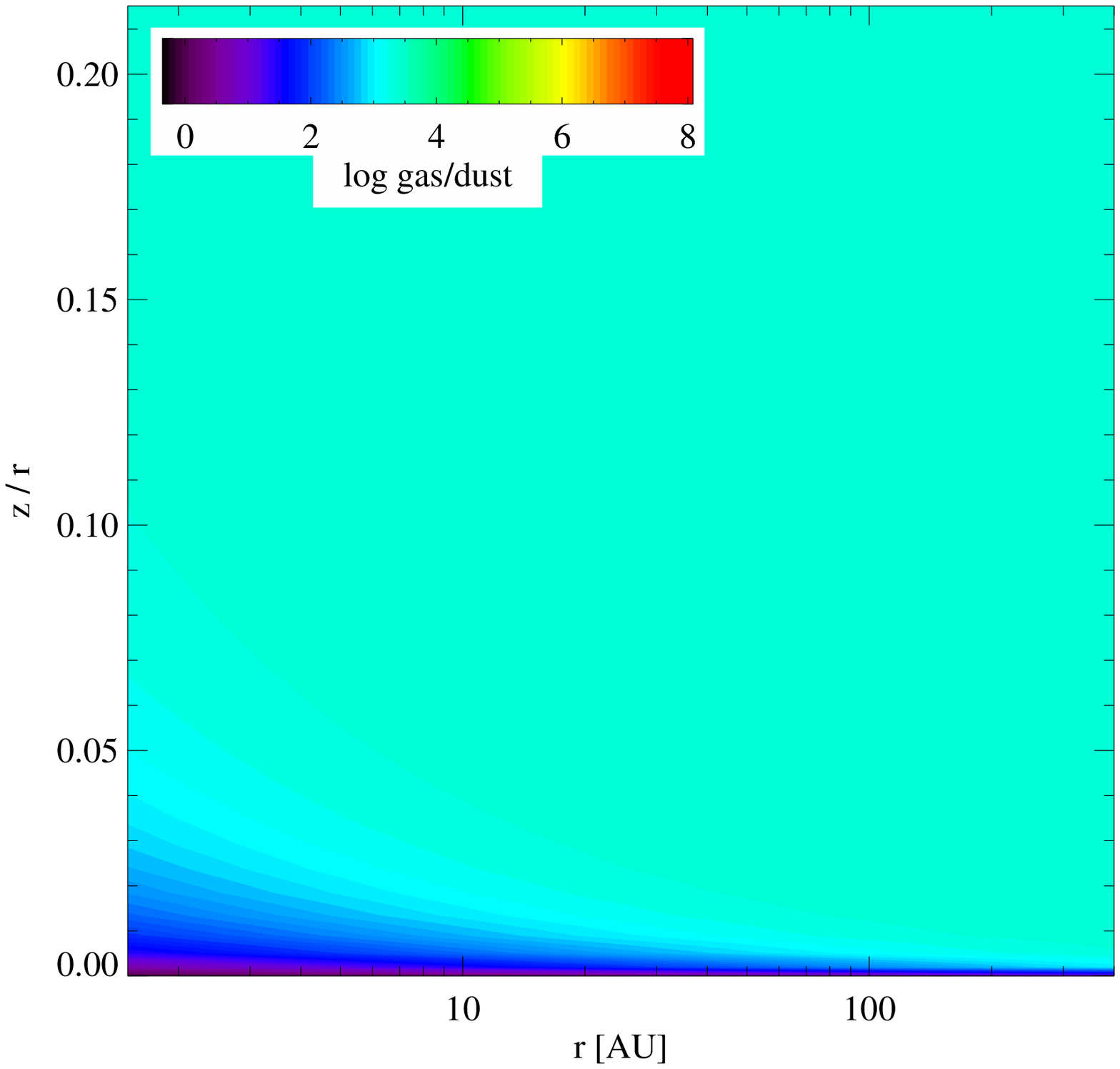}
 \includegraphics[scale=0.5]{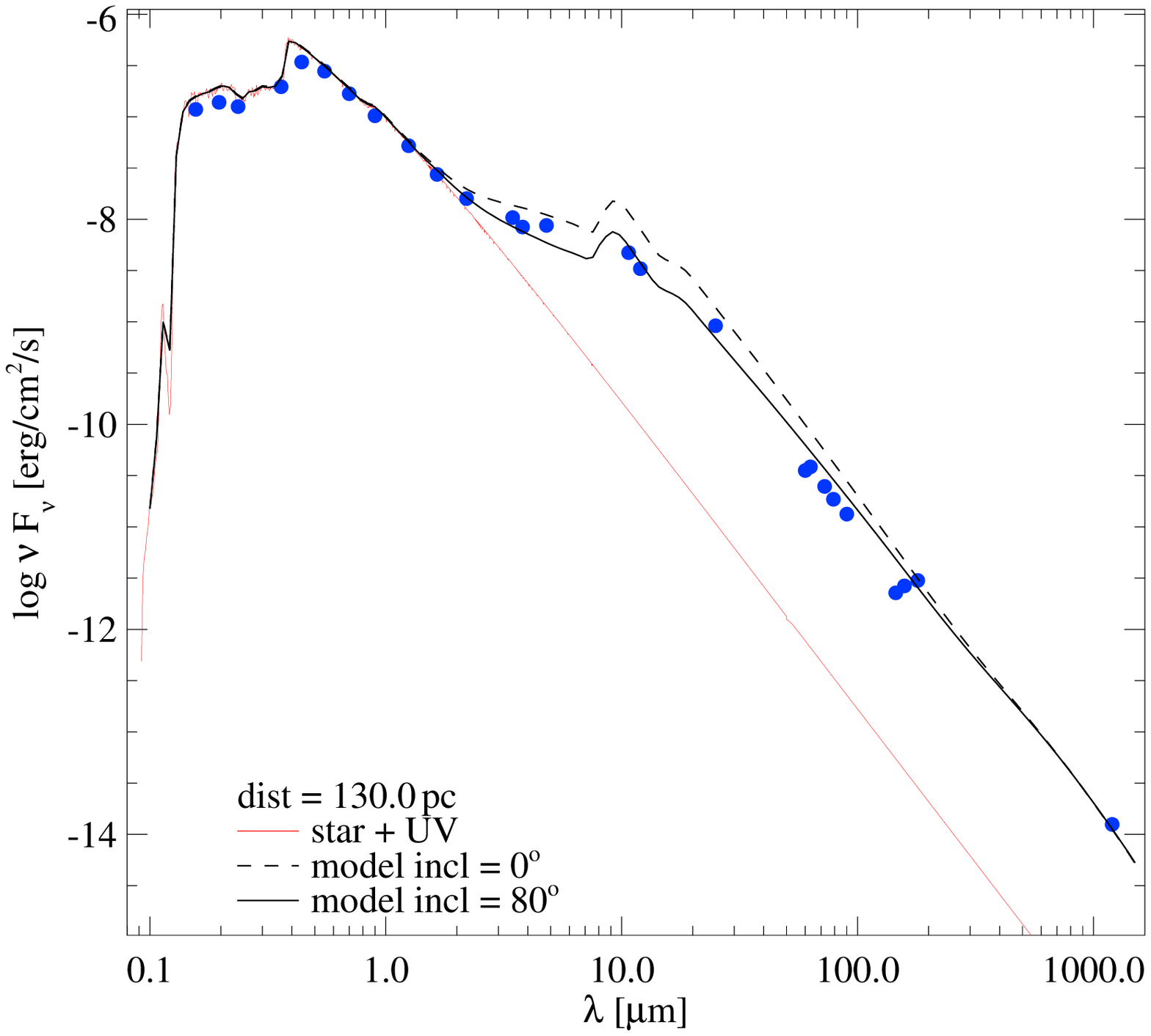}
 \includegraphics[scale=0.5]{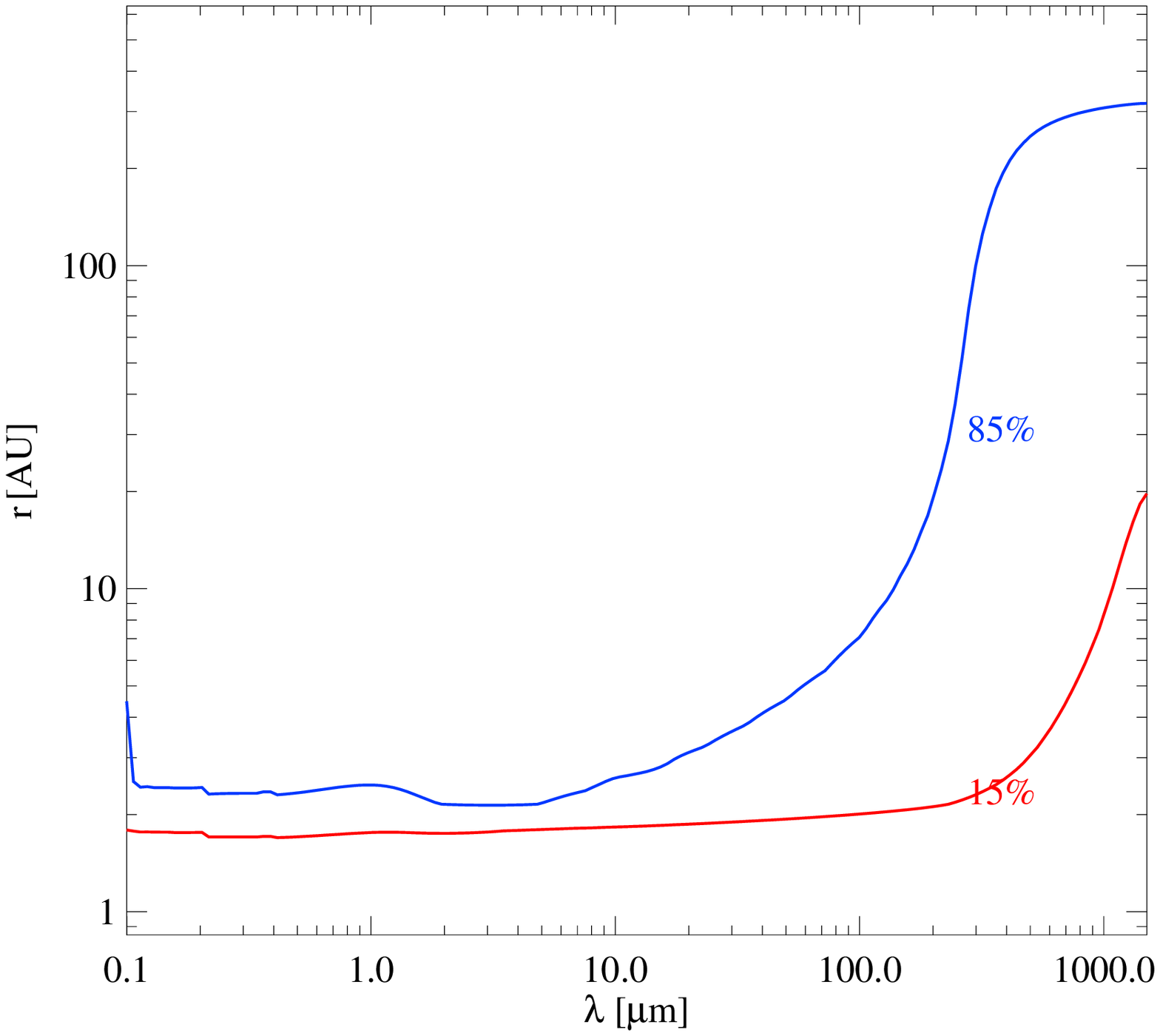}
 \caption{Disc gas density, gas-to-dust ratio, SED, and localization
   of the continuum emission for the fixed vertical structure disc
   model. The disc is gas-rich above the mid-plane and gas-poor in the
   mid-plane. For clarity a couple of observed photometric points are
   not shown in the SED. The lower-right panel shows the cumulative
   continuum emission for 15\% and 85\% for each wavelength. The
   continuum emission for wavelengths greater than 500$\mu$m comes
   from the outer disc ($R>$100~AU).}
  \label{fig_fixed_structure}            
\end{figure*}      
\begin{figure*}[!ht]
\centering
 \includegraphics[scale=0.5]{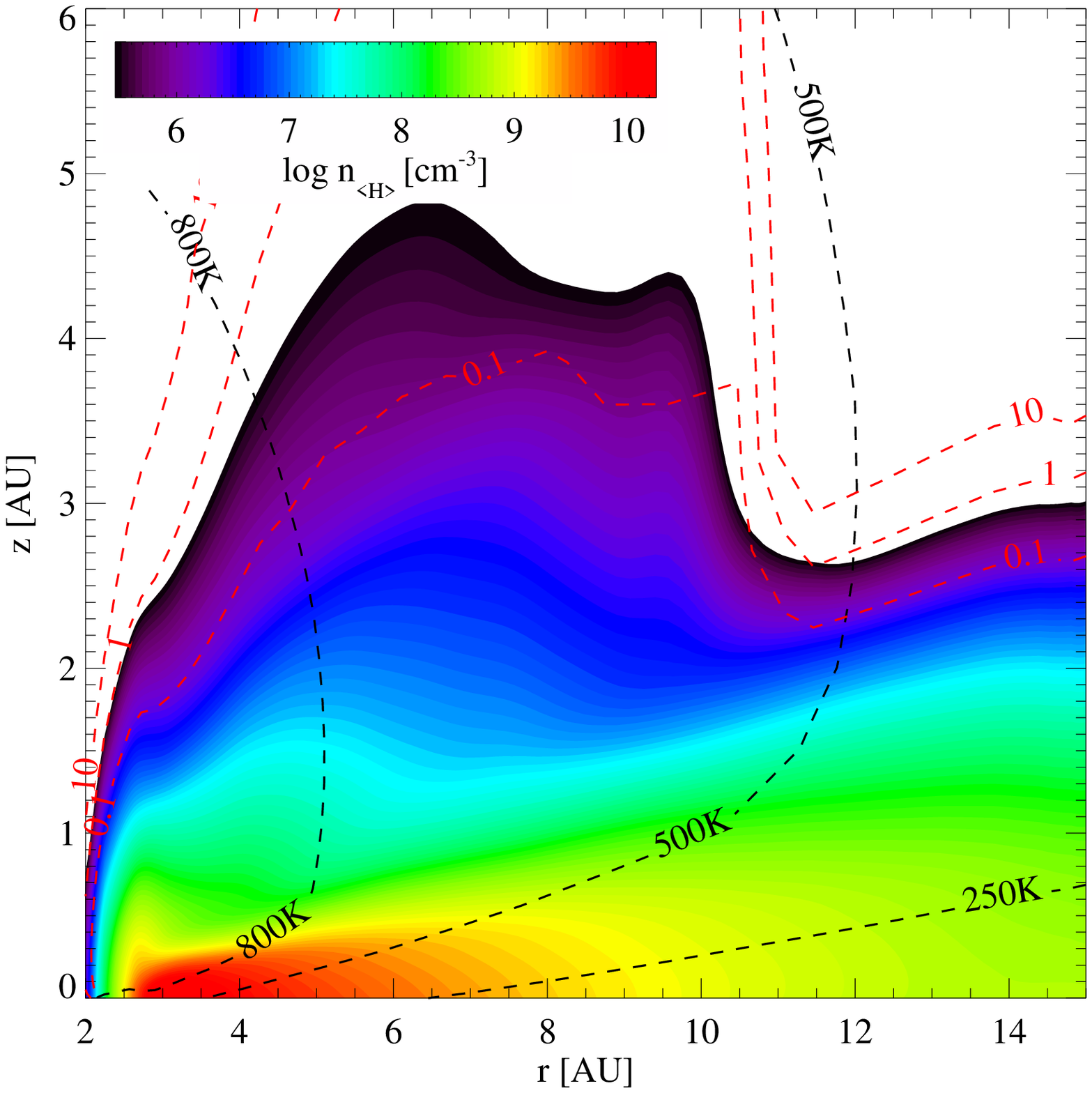}
 \hfill
 \includegraphics[scale=0.5]{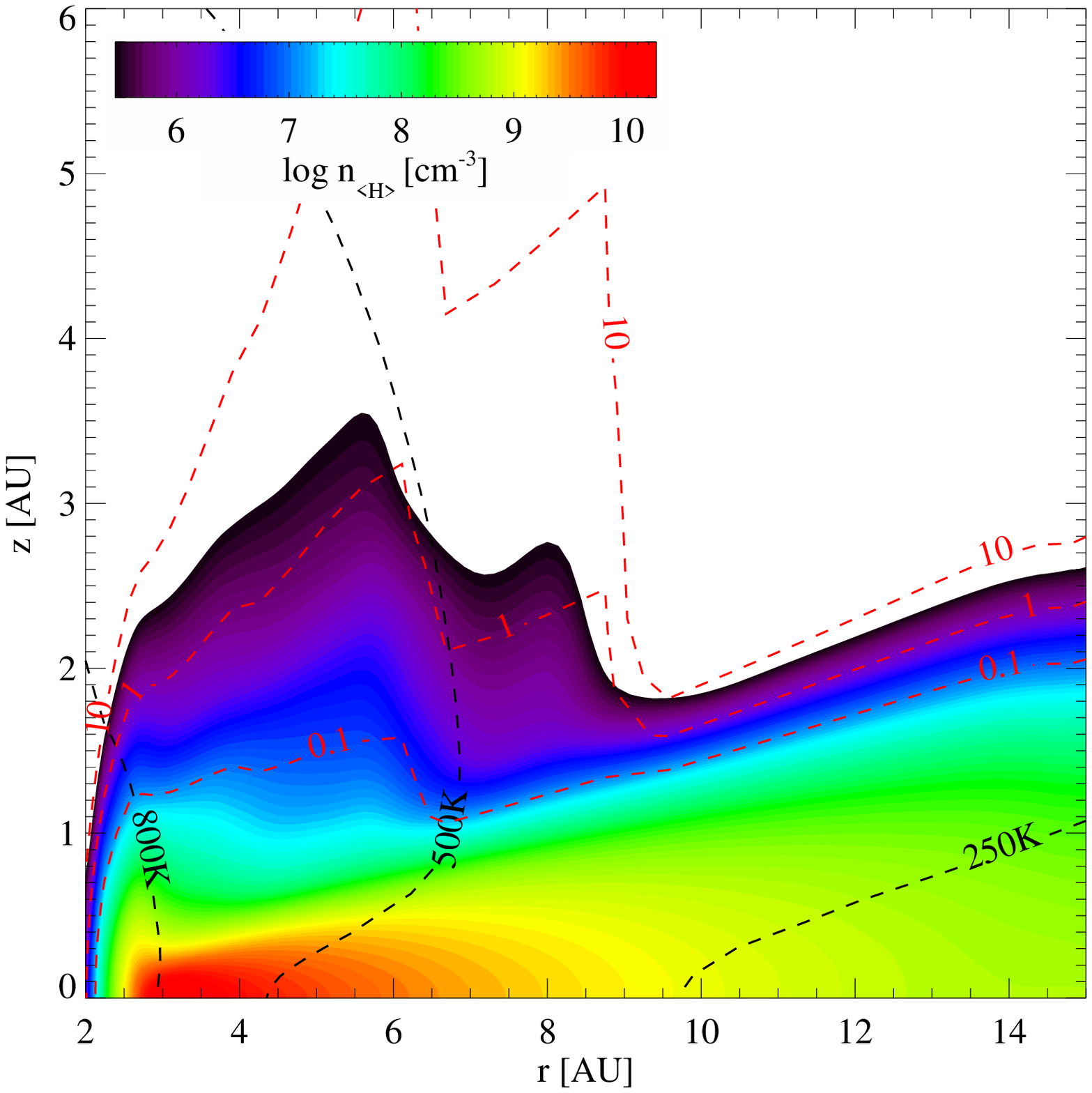}
 \caption{Disc gas density structure from hydrostatic models 1 and 2
   ($R_{\mathrm {out}}$=2 AU, $M_{\mathrm{disc}}$=5$\times$10$^{-6}$
   M$_\odot$) between 2 and 15~AU. The left panel corresponds to the
   model with dust settling (model 1), the right panel to the model
   without dust settling (model 2). The dust temperature contours are
   overplotted in black. The red contours correspond to the Stokes
   numbers ($S_{t}$=0.1, 1, and 10 contour values).}
  \label{fig_hydrostatic_density}          
\end{figure*}    

\begin{figure*}[!ht]
\centering
 \includegraphics[scale=0.5]{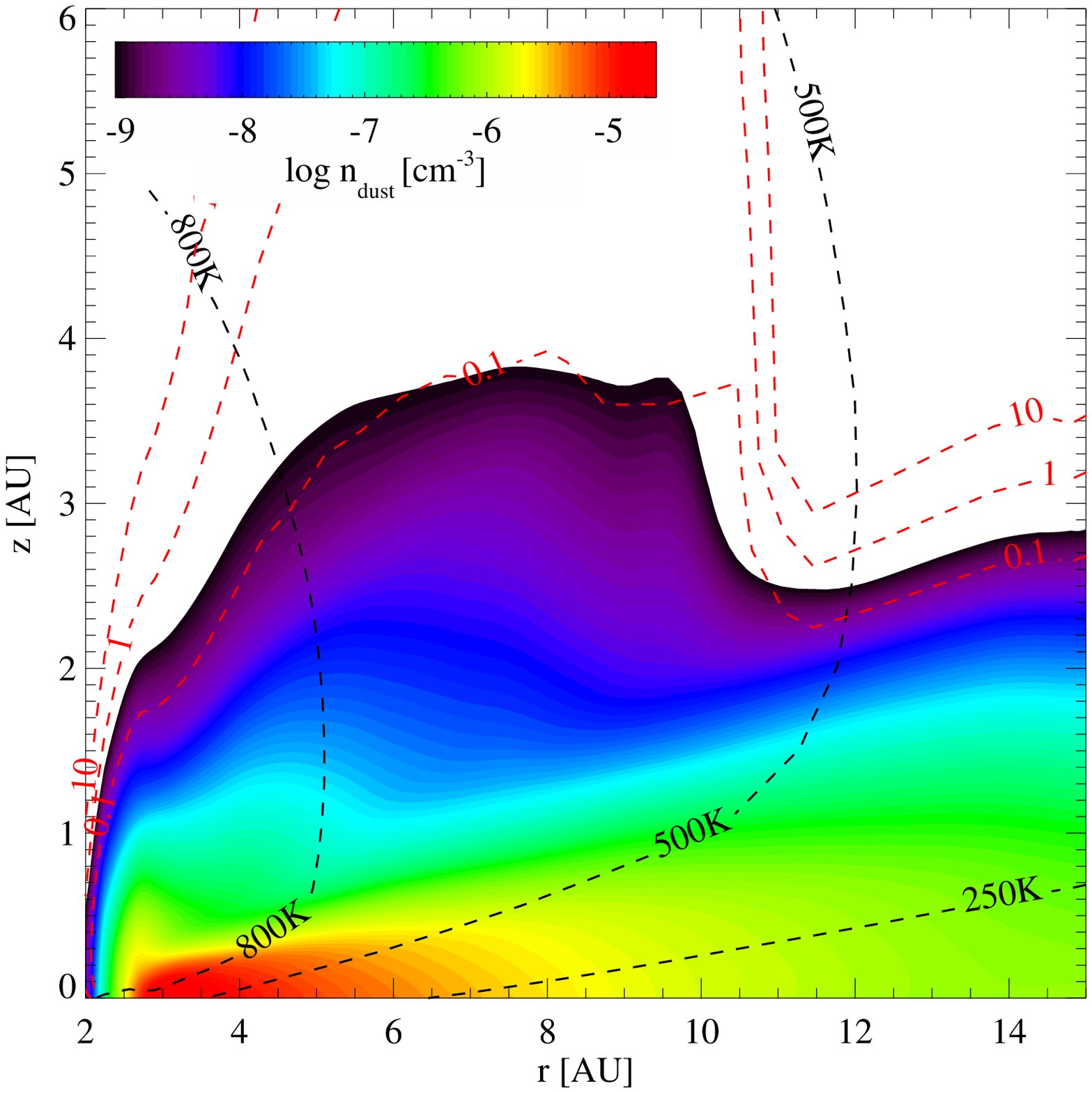}  
 \hfill  
 \includegraphics[scale=0.5]{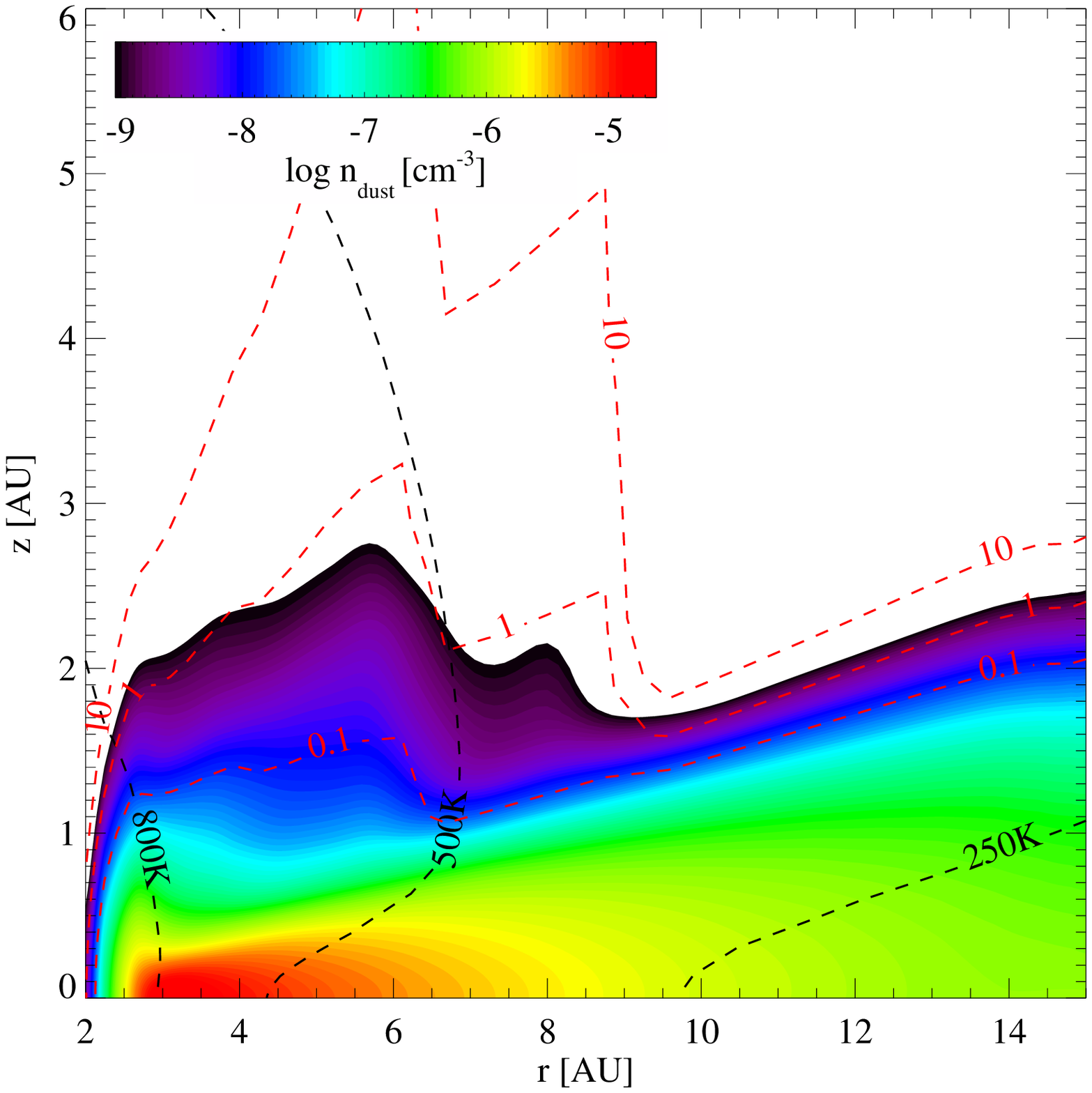}    
 \caption{Disc dust density structure from hydrostatic models 1 and 2
   ($R_{\mathrm {out}}$=2 AU, $M_{\mathrm{disc}}$=5$\times$10$^{-6}$
   M$_\odot$) between 2 and 15~AU. The left panel corresponds to the
   model with dust settling (model 1), the right panel to the model
   without dust settling (model 2). The dust temperature contours are
   overplotted in black.  The Stokes parameters are overplotted in red
   ($S_{t}$=0.1, 1, and 10 contour values).}
  \label{fig_hydrostatic_dust_structure}          
\end{figure*}    
\begin{figure}[!ht]
\centering
\resizebox{\hsize}{!}{\includegraphics[angle=0]{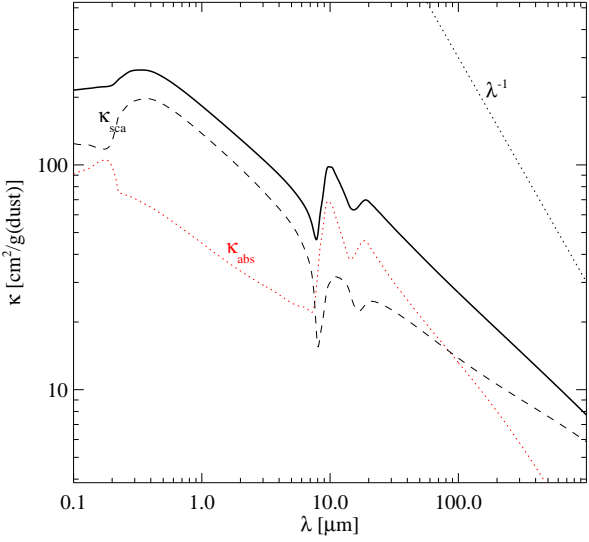}}
\caption{Grain mass absorption ($\kappa_{\mathrm{abs}}$ in red dotted-
  line), scattering ($\kappa_{\mathrm{sca}}$ in black dashed-line),
  and extinction ($\kappa$ in black full line) opacity for the adopted
  dust grain properties listed in Table~\ref{tab_DiscParameters}. The
  slope in the upper-right corner shows a hypothetical opacity with
  $\kappa\sim \lambda^{-1}$.}
\label{fig_kappa}          
\end{figure}  
\begin{figure}[!ht]
\centering
\resizebox{\hsize}{!}{\includegraphics[angle=0]{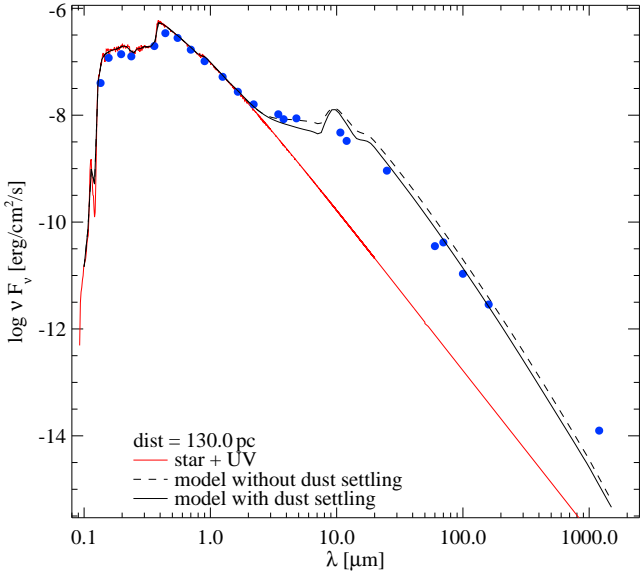}}
\caption{Fit to the SED for model 1 with and without dust settling. The four far-IR photometric points are {\it Herschel-PACS} observations. The 1.2~mm photometry was obtained with {\it MAMBO} at the {\it IRAM30m} telescope.} 
  \label{fig_hydrostatic_SED}          
\end{figure}    
\begin{figure}[!ht]
\centering
\resizebox{\hsize}{!}{\includegraphics[angle=0]{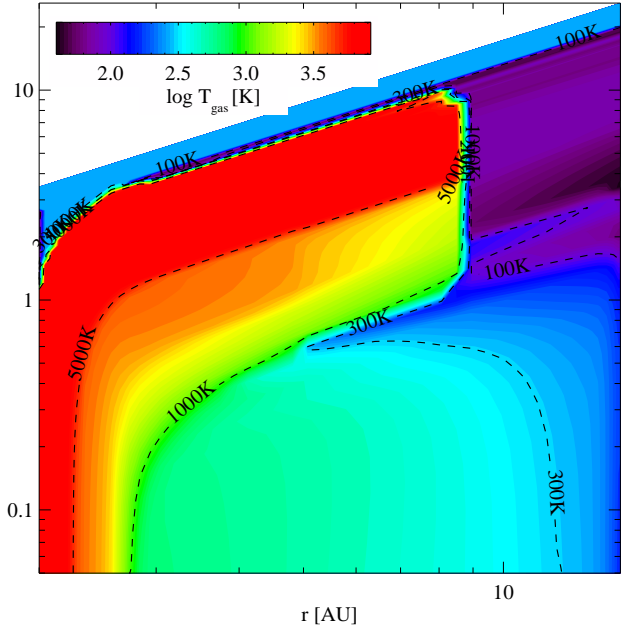}}
\caption{Gas temperature structure for model 1. The gas temperature is
  higher than 250~K in the inner compact disc.}
  \label{fig_hydrostatic_Tgas}          
\end{figure}    
\begin{figure}[!ht]
\centering
\resizebox{\hsize}{!}{\includegraphics[angle=0]{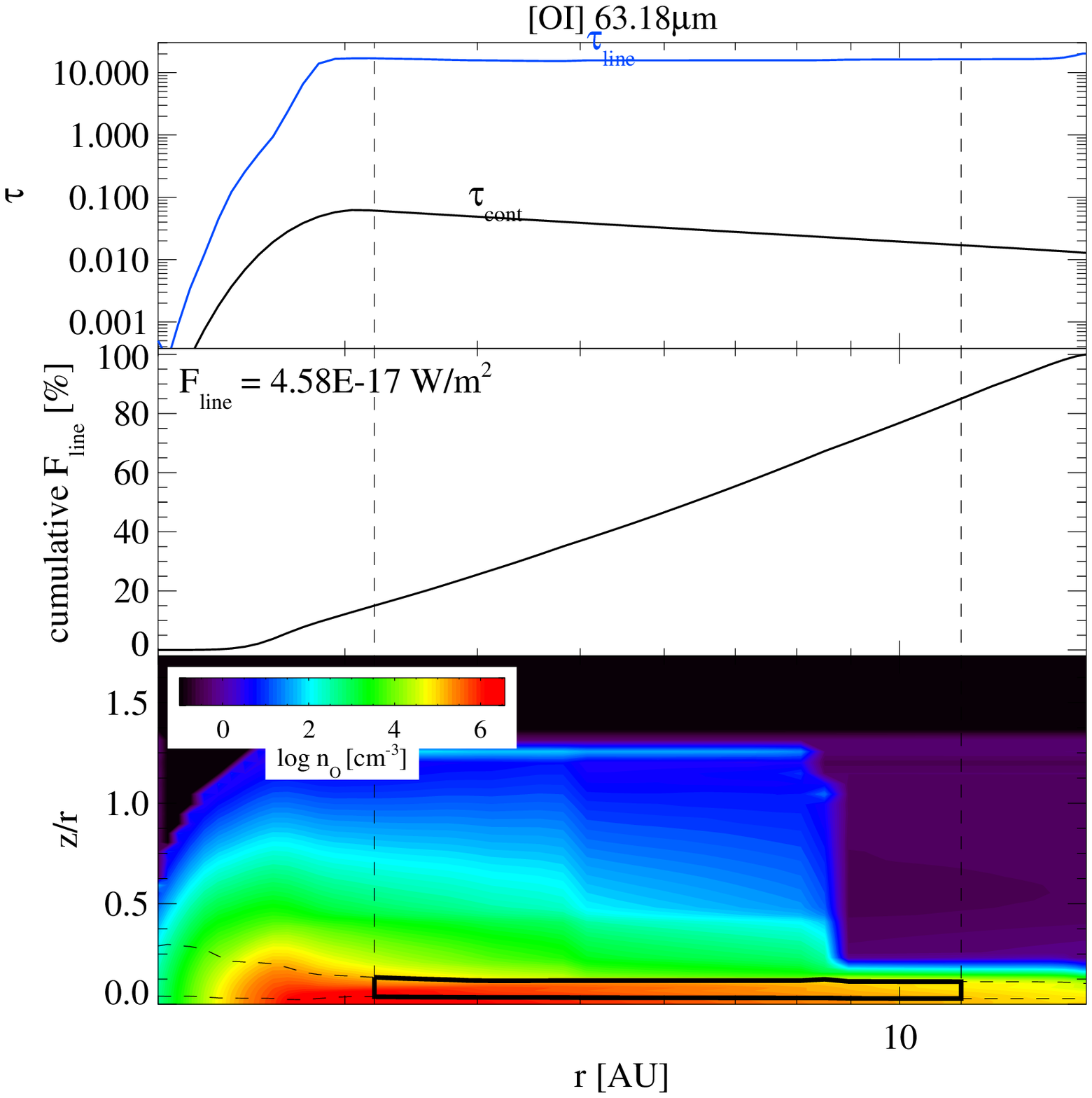}}
\caption{\OIfs\ emission from the disc for model 1. The upper panel
  shows the continuum at 63 microns and the line vertical optical
  depth. The middle panel shows the cumulative [OI] 63 micron emission
  from the inner to the outer disc. The lower panel exhibits the
  atomic oxygen density distribution in the disc. The contour encloses
  the 70\% $\times$ 70\% emission area.}
  \label{fig_hydrostatic_OI63_cumulative}          
\end{figure}    

We detected the spatially unresolved (on a scale of 10$"$) emission at
1.2mm with a flux of 5 $\pm$ 1 mJy (1 $\sigma$ error) towards
\object{51 Oph} using the MAMBO bolometer at the {\it IRAM 30m}
telescope, consistent with the upper limit of 42 mJy obtained at 1.1mm
by \citet{Sylvester1996MNRAS.279..915S}. The continuum was detected by
{\it PACS} in all settings. The total SED from the optical to the
millimeter is shown in Fig.~\ref{fig_herschel_spectra}.
\citet{Fajardo-Acosta1993ApJ...417L..33F} detected the 10~$\mu$m
silicate feature at low spectral resolution. The feature was later
confirmed by \citet{{vandenAncker2001A&A...369L..17V}}. The millimeter
flux translates into a dust grain (with radii $<$ 1mm) mass of
(1-2)~$\times$ 10$^{-6}$ M$_\odot$ assuming $T_{\mathrm{dust}}$= 50 K
and the standard dust opacity $\kappa_\nu = 2.0
(\nu/\nu_0)^{\beta}$\,g$^{-1}$\,cm$^{2}$ where $\nu_0=230.769$\,GHz
and $\beta=0.6$ \citep{Beckwith1991ApJ...381..250B}. This derived
value is an upper limit to the dust mass since we cannot estimate the
contribution from free-free emission at 1.2mm due to the lack of data
at longer wavelengths. Alternatively, pebbles with size greater than 1
cm will emit as blackbodies in the millimeter range. Assuming a
gas-to-dust mass ratio of 100, the disc gas mass is (1-2)~$\times$
10$^{-4}$ M$_\odot$.

We complemented the SED with archival photometric data and the {\it
  Spitzer-IRS} spectrum (see Fig.~\ref{fig_sed_modified_bb}). We
fitted the SED with a Kurucz stellar model in addition to a modified
blackbody of the form
\begin{equation}
\nu F_\nu = B_\nu(T)(\lambda/\lambda_0)^{-\beta-1},
\end{equation}
where $B_\nu(T)$ is the Planck function at the temperature $T$. In the
formula, we assumed that the grains will emit with $\kappa=\kappa_0$
\,g$^{-1}$\,cm$^{-}$ for $\lambda < \lambda_0$ and with $\kappa_\nu =
\kappa_0 (\lambda/\lambda_0)^{-\beta}$\,g$^{-1}$\,cm$^{2}$ for
$\lambda \ge \lambda_0$. The value of $\lambda_0$ is set by the
criterion $\lambda_0>2\pi <a>$, where $<a>$ is the mean grain
radius. We adopted $\lambda_0$ = 100 $\mu$m ($<a>$= 16 $\mu$m). The
fits to the SED for three values of $\beta$ ($\beta$=1, 0, -0.3) are
shown in Fig.~\ref{fig_sed_modified_bb}.  A value for $\beta$ of 1 is
typical for grains found in protoplanetary discs
\citep{Draine2006ApJ...636.1114D}. A pure blackbody emission has
$\beta$=0. Alternatively, a grain size powerlaw distribution with an
index $p=3$ will result in a quasi-constant $\kappa$ with wavelength
\citep{Draine2006ApJ...636.1114D}. The model with the standard value
for protoplanetary discs $\beta$=1 fails to match the millimeter flux.
A pure blackbody emission also underpredicts the millimeter flux, but
the prediction is still consistent with the observations within
3~$\sigma$. The best fit is obtained with $\beta$=-0.3. A value
  for $\beta$ of -0.3 is unphysical and points to the emission by an
  extra component, either a cold outer disc/envelope or free-free
  emission.

  In addition, far-IR spectral line observations were obtained using
  the Herschel Space Observatory. The only detected line is the \OIfs\
  emission at 63 microns. The \OIfs\ emission is confined to the
  central spaxels. This line is always the strongest in the two \OIfs\
  and \CIIfs\ surveys of the {\it GASPS} project
  \citep{Meeus2012A&A...544A..78M}. No high-$J$ CO or water lines were
  detected. The line flux and three sigma upper limits are summarized
  in Table~\ref{table_results}. Finally, we show the archival
  H$\alpha$ line profile in Fig.~\ref{fig_Halpha_spectrum}. The
  profile shows a double-peaked structure on the top of the stellar
  H$\alpha$ absorption. The origin of this profile is discussed in
  Sec.~\ref{modelling}.

\begin{table}[!htb]
\centering
\caption{Lines and photometric data obtained by {\em Herschel-PACS}. The errors are 1$\sigma$ and the upper limits are 3$\sigma$. The calibration error adds an extra $\sim$~10\% uncertainty.}
\label{table_results} 
\begin{tabular}{lr@{.}lr@{~$\pm$~}lr@{}r@{.}l}   
  \hline 
  \noalign{\smallskip}   
  \multicolumn{1}{c}{Line} & \multicolumn{2}{c}{$\lambda$} & \multicolumn{2}{c}{Cont. flux} & \multicolumn{3}{c}{Line flux}\\
  & \multicolumn{2}{c}{($\mu$m)}  &  \multicolumn{2}{c}{(Jy)}& \multicolumn{3}{c}{(10$^{-18}$ W m$^{-2}$)}\\    
  \noalign{\smallskip}   
  \hline
  \noalign{\smallskip} 
  \OIfs\  $^3$P$_1 \rightarrow ^3$P$_2$ & 63&183 & 1.09&0.19 & & 53&3~$\pm$~2.5\\
  \OIfs $^3$P$_0 \rightarrow ^3$P$_1$ & 145&525 &  0.11&0.10 & $<$ & 5&1  \\
  \CIIfs\  $^2$P$_{3/2} \rightarrow ^2$P$_{1/2}$ &  157&74 & 0.07&0.14  & $<$& 6&8 \\
  \noalign{\smallskip} 
  \hline
  \noalign{\smallskip} 
  CO $J$=36 $\rightarrow$ 35   & 72  & 85 &  0.71& 0.20 & $<$ &  11&6\\
  CO $J$=33 $\rightarrow$ 32   & 76  & 36 &  0.52  & 0.28 & $<$ &  16&1\\
  CO $J$=29 $\rightarrow$ 28  & 90&16 & 0.50&0.19 & $<$ &   8&0\\
  CO $J$=18 $\rightarrow$ 17  & 144&78 & 0.12&0.07 & $<$ &  4&2\\
  \noalign{\smallskip} 
  \hline
  \noalign{\smallskip} 
  OH $1/2-3/2$ & 79& 11 & 0.54& 0.19 & $<$ & 11 & 4\\
  OH $1/2-3/2$ & 79 & 18 & 0.50 & 0.19 & $<$ & 11 & 3\\
  \noalign{\smallskip} 
  \hline
  \noalign{\smallskip} 
  o-H$_2$O $8_{18}-7_{07}$   & 63& 32 & 1.08    & 0.15 &$<$ & 8&3 \\
  o-H$_2$O $7_{07}-6_{16}$   & 71&94  & 0.71 & 0.23 & $<$ & 16&1\\ 
  p-H$_2$O $3_{22}-2_{11}$   & 89  & 99 &  0.45&0.24  & $<$ &   10&3\\
  p-H$_2$O $4_{13}-3_{22}$   & 144 &51 & 0.12&0.08  & $<$ & 4&3\\
  p-H$_2$O $3_{31}-4_{04}$   & 158 & 31 & 0.09&0.15 & $<$ & 7&3\\
  o-H$_2$O $2_{12}-1_{01}$   & 179   & 52   & 0.24&0.19 & $<$ & 7&0\\
  o-H$_2$O $2_{21}-2_{12}$   & 180   & 42   & 0.13&0.16 & $<$ & 5&7\\
  \noalign{\smallskip} 
  \hline
  \noalign{\smallskip} 
  CH$^+$ $J=5-4$ & 72 & 14 & 0.64 & 0.3 &$<$ & 19&3\\
  CH$^+$ $J=4-3$ & 90 & 01 & 0.45 & 0.24 & $<$ & 10&3\\
\noalign{\smallskip} 
\hline
                  \noalign{\smallskip}   

\multicolumn{8}{c}{{\it Herschel-PACS} continuum data}\\
                  \noalign{\smallskip}     
\multicolumn{1}{c}{...}  & 70  &0 & 0.873  & 0.023 & \multicolumn{1}{c}{...}\\
\multicolumn{1}{c}{...}  & 100  &0 & 0.347  &0.010 & \multicolumn{1}{c}{...}\\
\multicolumn{1}{c}{...}   & 160 &0 & 0.090  &0.006& \multicolumn{1}{c}{...}\\
\hline
\end{tabular}
\end{table}
\normalsize 
\section{SED and gas modelling}\label{modelling}

We modelled the continuum and gas emission with the codes {\sc MCFOST}
and {\sc ProDiMo}.  {\sc MCFOST} is a continuum and NLTE line Monte
Carlo radiative transfer code that can handle 2D and 3D geometries
\citep{Pinte2006A&A...459..797P,Pinte2009A&A...498..967P}. {\sc
ProDiMo} is a radiative chemo-physical code that solves the continuum
radiative transfer, the gas heating and cooling balance, the gas
chemistry, and the disc vertical hydrostatic structure
self-consistently with the gas pressure
\citep{Woitke2009A&A...501..383W}.  New features of the codes are
described in \citet{Kamp2010A&A...510A..18K},
\citet{Thi2011MNRAS.412..711T}, and
\citet{Woitke2011A&A...534A..44W}. The 2D continuum radiative
  transfer module in {\sc ProDiMo} and the code {\sc MCFOST} has been
  benchmarked against other codes \citep{Pinte2009A&A...498..967P}.

  The modelling strategy is twofold. First we assumed a fixed vertical
  structure disc with dust settling. The disc extends from 1.5~AU to
  400~AU as suggested by the model of
  \citet{Stark2009ApJ...703.1188S}. We employed an automatic fit
  procedure based on the Simplex optimization algorithm. The free
  parameters of the model (model a) are the total disc gas+dust mass,
  the surface density profile index, the gas-to-dust ratio, the disc
  gas scale-height, and the flaring index. In model b, the free
  parameters are the same except for the gas-to-dust ratio, which is
  fixed at 100. The dust scale height including settling is
\begin{equation}
H'(a,r)=H(r)\mathrm{max}[1,a/a_{\mathrm{s}}]^{-s/2},
\end{equation}
where $H(r)$ is the gas scale height, and $s$ and $a_{\mathrm{s}}$ are
two fixed parameters: $s$ is the power-law index and $a_{\mathrm{s}}$
is the minimum radius for a grain to be affected by settling. The
settling law is described in details in
\citet{Woitke2009A&A...501..383W} and has previously been used in
modelling the \object{HD~163296} disc
\citep{Tilling2012A&A...538A..20T}. We adopted a minimum dust radius
for settling $a_{\mathrm{s}}$ of 0.5 $\mu$m and $s$=1. The disc
density and gas-to-dust ratio structure are shown in
Fig.~\ref{fig_fixed_structure}.

\begin{center}
\begin{table}[!ht]
  \caption{Common disc parameters. The parameters are defined in \citet{Woitke2009A&A...501..383W}.}\label{tab_DiscParameters}
		\begin{tabular}{lll}
                  \hline
                  \noalign{\smallskip}   
                  Stellar mass & $M_*$ & 3.8~M$_\odot$ \\ 
                  Stellar luminosity &$L_*$ & 260~L$_\odot$ \\ 
                  Effective temperature & $T_{\mathrm {eff}}$ & 10000~K\\
                  Distance        & $d$ & 130~pc\\
                  Disc inclination & $i$ & 80~\degr\ (0~\degr face-on)\\
                  Disc inner radius & $R_{\mathrm {in}}$  & 1.5 or 2.0~AU \\
                  Disc outer radius & $R_{\mathrm {out}}$  & 15 or 400~AU\\
                  Vert. column density index & $\epsilon$ & 1 \\
                  Inner rim soft edge  & & on\\
                  Gas-to-dust mass ratio           &  $\delta$ & 100 \\
                  Dust grain material mass density & $\rho_{\mathrm{dust}}$ & 3.5 g cm$^{-3}$ \\
                  Minimum dust particle size       & $a_{\mathrm{min}}$ & 0.1 $\mu$m\\
                  Maximum dust particle size       & $a_{\mathrm{max}}$ & 5000 $\mu$m\\
                  Composition                      & & ISM silicate\\
                  Dust size distribution power law & $p$            & 3.5\\
                  H$_2$ cosmic ray ionization rate       & $\zeta_{\mathrm{CR}}$  & 1.7$\times$10 $^{-17}$ s$^{-1}$\\
                  ISM UV field w.r.t. Draine field  & $\chi$          & 1.0 \\
                  Abundance of PAHs relative to ISM & $f_{\rm PAH}$      & 10$^{-3}$\\
                  $\alpha$ viscosity parameter      & $\alpha$           &  0.0 \\
                  Turbulence width                  & $\delta v$         & 0.15~km~s$^{-1}$\\
                  \noalign{\smallskip}   
                  \hline
                  \noalign{\smallskip}   
                  Total disc mass & $M_{\mathrm{disc}}$ & 5$\times$10$^{-6}$-1$\times$10$^{-4}$ M$_\odot$ \\ 
                  Dust settling power law & $s$ & 0, 1.0\\
                  Minimum grain radius for settling & $a_{\mathrm{s}}$ & 0.5~$\mu$m\\
\hline
\end{tabular}  
\ \\ 
\end{table}
\end{center}
Second, we relaxed the assumption of fixed disc vertical structure by
letting the {\sc ProDiMo} code calculate the gas scale-height that
would be consistent with the local gas pressure as determined by the
gas heating and cooling balance.

The disc dust and gas parameters are gathered in
Table~\ref{tab_DiscParameters}.  The stellar parameters are taken from
\citet{Montesinos2009A&A...495..901M}. The input stellar spectrum is
taken from the Kurucz database of models.  The disc is seen close to
edge-on between 80\degr--90\degr\
\citep{Thi2005A&A...430L..61T,Berthoud2007ApJ...660..461B,Tatulli2008A&A...489.1151T}. The
optical photometry shows that the star is seen without strong
circumstellar extinction, therefore the dust disc has to be flat. On
the other hand, the gas disc should be more extended vertically such
that absorption lines can be detected
\citep{Grady1993ApJ...402L..61G,Lecavelier1997A&A...321L..39L,Roberge2002ApJ...568..343R}. Therefore
we tested disc models where the gas and dust are well mixed and where
the dust grains have settled towards the midplane.

The accretion rate of the \object{51 Oph} disc is disputed. Assuming a
magnetospheric origin, \citet{GarciaLopez2006A&A...459..837G} and
\citet{Brittain2007ApJ...659..685B} estimated a gas accretion rate of
(1-2)$\times$10$^{-7}$ M$_\odot$ yr$^{-1}$ from the hydrogen
Br$\gamma$ line. However, \citet{Mendigutia2011A&A...535A..99M} found
that the H$\alpha$ 10\% peak width for \object{51 Oph} is much lower
than the expected minimum value required by the magnetospheric
accretion model (Fig.~\ref{fig_Halpha_spectrum}). Therefore, they
argued that the inner disc of \object{51 Oph} is decoupled from the
star and that the mass accretion rate of 10$^{-7}$ M$_\odot$ yr$^{-1}$
is an upper limit. In a subsequent paper,
\citet{Mendigutia2012A&A...543A..59M} derived an empirical relation
between the disc mass $M_{\mathrm{disc}}$ in M$_\odot$ and the disc
mass accretion $\dot{M}_{\mathrm{acc}}$ in M$_\odot$ yr$^{-1}$ valid
for both T~Tauri and Herbig~Ae stars
\begin{equation}
  \log{\dot{M}_{\mathrm{acc}}}=1.1(\pm 0.3)\times \log{M_{\mathrm{disc}}}-5.0(\pm 0.5). 
\end{equation}
Using this relation and upper limit to the gas disc mass of
2$\times$10$^{-4}$ M$_\odot$ derived from the 1.2mm flux assuming a
gas-to-dust mass ratio of 100, we obtained an upper limit to the mass
accretion rate of $\sim$ 8.5$\times$10$^{-10}$ M$_\odot$ yr$^{-1}$. In
this paper, we modelled the disc around \object{51 Oph} as passive
($\alpha$ viscosity parameter is 0 in Table~\ref{tab_DiscParameters}).
The H$\alpha$ emission most likely does not trace magnetospheric
accretion. The origin of the double-peaked H$\alpha$ emission profile
can be ascribed to a stellar/disc wind
\citep{Kraus2008A&A...489.1157K}. A low turbulent width of 0.15 km
s$^{-1}$ was assumed.

The inner disc radius is consistent with the location of the
sublimation radius for a grain size distribution that follows a
power-law with index 3.5 ($a_{\mathrm{min}}$=0.1~$\mu$m and
$a_{\mathrm{max}}$=2000~$\mu$m). The dust grains are composed of pure
amorphous silicate (Astronomical silicates,
\citealt{Laor1993ApJ...402..441L}). The choice of amorphous silicate
is motivated by the absence of strong crystalline silicate features in
the infrared \citep{Keller2008ApJ...684..411K}. The dust opacities
computed using the Mie theory for compact spherical particles are
shown in Fig.~\ref{fig_kappa}. Except from 10 and 100 $\mu$m, the
opacity is dominated by the scattering term.

Even a perfect fit to the SED alone does not allow one to constrain
the disc size, which is a key parameter in many physical phenomena:
dust coagulation rate, dust-gas interaction,
etc \citep{Armitage2010apf..book.....A}. One of the main uncertainties
is the disc outer radius. We adopted two values for the outer disc: a
small disc with $R_{\mathrm {out}}$=15~AU and an large disc with
$R_{\mathrm {out}}$=400~AU.  The high value for the outer radius at
400 AU was motivated by the model of \citet{Stark2009ApJ...703.1188S}
while a compact disc is suggested by the drop in the continuum flux at
wavelengths beyond 20 micron. We also modelled various discs with
a  sub-standard gas-to-dust ratio (models 9 to 16).

The \OI\ emission at 63.18 $\mu$m probes only relatively warm and
dense gases because of its upper energy level at 227~K and a critical
density of $\sim$4.7$\times$10$^{5}$ cm$^{-3}$. In the absence of
detected lines that probe cold gases, we cannot constrain the
gaseous content beyond 10-20~AU well.

\begin{table}[!ht]
\begin{center}
  \caption{Best-fit model parameters for the fixed vertical disc structure model. \label{model_fixed_structure_description}}
\begin{tabular}{lll}
  \hline
  \hline
  \noalign{\smallskip}  
  Total disc gas mass                 & $M_{\mathrm{gas+dust}}$ (M$_{\odot}$)       & 1.1$\times$10$^{-4}$ \\
  Gas-to-dust mass ratio        &       & 29, 100\\
  Abundance of PAHs relative to ISM & $f_{\rm PAH}$      & 10$^{-3}$ \\
  Surface density profile index & $\epsilon$  & 0.4 \\
  Reference gas scale height               & $H_0$ (AU) &  0.14 \\
  Reference radius       & $R_{\mathrm{ref}}$ (AU)       & 5.0 \\
  Flaring index                 &  $\gamma$      & 0.55  \\
  \hline
\end{tabular}  
\ \\ 
\end{center}
\end{table}


\begin{table*}[!ht]
\begin{center}
  \caption{Model parameters and line fluxes with 3~$\sigma$ upper
    limits. \label{model_fluxes}}
\begin{tabular}{llllllll}
  \hline
  \hline
  \noalign{\smallskip}   
  Model & $R_{\mathrm {out}}$ & $M_{\mathrm{disc}}$ & gas-to-dust  & settling & \OIfs\  63 $\mu$m & \OIfs\  145 $\mu$m & \CIIfs\  157 $\mu$m\\
  &    (AU)         &   (M$_\odot$ )   &  ratio          & \multicolumn{3}{c}{(10$^{-18}$ W m$^{-2}$)}\\
  \noalign{\smallskip}   
  \hline
  \noalign{\smallskip} 
  & & & &  & \multicolumn{3}{c}{Observations}\\
  \noalign{\smallskip} 
    &                 &                   &    &   & 53.3~$\pm$~2.5 & $<$5.1 & $<$6.8\\
  \noalign{\smallskip}   
  \hline
  \noalign{\smallskip} 
 & & & &  & \multicolumn{3}{c}{Fixed vertical structure model}\\  
  a    &   400           &  1.0$\times$10$^{-4}$ & 29   & yes & 35.8 & 1.3 & 0.17 \\
  b    &   400           &  1.0$\times$10$^{-4}$ & 100  & yes & 65.4 & 3.6 & 0.02 \\
  \noalign{\smallskip}   
  \hline
  \noalign{\smallskip}   
 & & & &  & \multicolumn{3}{c}{Vertical hydrostatic models}\\
  1     &    15           & 5$\times$10$^{-6}$ & 100  & yes & 47.2  & 2.8 & 0.02 \\  
  2     &    15           & 5$\times$10$^{-6}$ & 100  & no  & 49.4  & 2.9 & 0.03 \\
  3     &   400           & 5$\times$10$^{-6}$ & 100  & yes & 22.2  & 0.6 & 1.2  \\ 
  4     &   400           & 5$\times$10$^{-6}$ & 100  & no  & 22.2  & 0.6 & 1.2  \\ 
  5     &   400           & 5$\times$10$^{-5}$ & 100  & yes & 65.3  & 2.8 & 1.1  \\
  6     &   400           & 5$\times$10$^{-5}$ & 100  & no  & 64.9  & 2.7 & 1.0  \\
  7     &   400           & 1$\times$10$^{-4}$ & 100  & yes & 94.6  & 4.2 & 1.2  \\
  8     &   400           & 1$\times$10$^{-4}$ & 100  & no  & 93.7  & 4.2 & 1.2  \\ 
  9     &    15           & 5$\times$10$^{-7}$ &  10  & yes & 38.2  & 1.1 & 0.03 \\
 10     &    15           & 5$\times$10$^{-7}$ &  10  & no  & 40.8  & 1.2 & 0.03 \\
 11     &    15           & 5$\times$10$^{-8}$ &   1  & yes & 5.3   & 0.1 & 0.005 \\
 12     &    15           & 5$\times$10$^{-8}$ &   1  & no  & 5.3   & 0.1 & 0.005 \\
 13     &   400           & 5$\times$10$^{-5}$ &  10  & yes & 144.7 & 4.2 & 9.8   \\
 14     &   400           & 5$\times$10$^{-5}$ &  10  & no  & 134.6 & 3.9 & 8.4   \\
 15     &   400           & 5$\times$10$^{-6}$ &   1  & yes & 39.5  & 1.0 & 2.5  \\
 16     &   400           & 5$\times$10$^{-6}$ &   1  & no  & 39.5  & 1.0 & 2.5  \\
  \noalign{\smallskip}   
  \hline
\end{tabular}  
\end{center}
\ \\ 
\end{table*}
One of the major unknown in the gas line modelling is the amount of
PAHs in the disc. The {\it Spitzer-IRS} spectrum in
Fig.~\ref{fig_sed_modified_bb} does not show the typical IR features
of PAHs \citep{Keller2008ApJ...684..411K,Stark2009ApJ...703.1188S}. We
ran the models with the PAH abundance depleted with respect to the
interstellar value of 3~$\times$~10$^{-7}$ by a factor 10$^{-3}$.

The second source of uncertainties are the elemental abundances
relative to H-nuclei. In young protoplanetary discs where the gas
comes for the collapse of a prestellar core, the elemental abundances
are the same as for molecular clouds. For debris discs, the gas
comes from the evaporation of icy planetesimals and the gas is
enriched in elements heavier than hydrogen and helium. The
evolutionary state of \object{51~Oph} is disputed. The presence of
significant amount of hot CO as evidenced by the bandhead emission, of
water vapour, and of carbon dioxide support the hypothesis that the
oxygen and carbon elemental abundance with respect to hydrogen are
close to the molecular cloud values.

The chemical network was limited to 71 gas and ice species including
formation of H$_2$ on grain surfaces and fast neutral reaction
involving vibrationally excited H$_2$. Most of the reaction rates are
taken from the UMIST database \citep{Woodall2007A&A...466.1197W}. We
adopted a standard value of 1.7$\times$10 $^{-17}$ s$^{-1}$ for the
cosmic-ray ionization rate.

The code takes cooling by water lines from the near-IR to the far-IR
into account. The collisional data were taken from the {\it
  Leiden-Lambda} database \citep{Schoier2005A&A...432..369S}. The
original references for the experimental or theoretical rates are
given in appendix~\ref{collisional_rates}.  The observed \OIfs\ flux
is best reproduced by a compact (15 AU) disc with masse of
5$\times$10$^{-6}$ M$_\odot$. The disc gas-to-dust mass ratio is
consistent with the interstellar value of 100.  All other modelled
fluxes are below the 3 sigma upper limits.

We ran two groups of vertical hydrostatic models: one with a compact
disc and one with a large disc. The parameters of the models are
listed in Table~\ref{model_fluxes}.

\section{Model results}\label{model_results}

We first present the results from the Simplex automatic fitting using
the fixed vertical structure disc model. The fitting procedure leads
to an extremely flat disc for the gas ($H/r$=2\%). The dust disc is
even flatter because of settling (upper panels of
Fig.~\ref{fig_fixed_structure}).  The best model has a gas mass of
1$\times$10$^{-4}$ M$_\odot$ and a dust mass of 1$\times$10$^{-6}$
M$_\odot$, similar to the values derived from the simple analysis of
the millimeter emission. The best model fits to the SED and the
continuum emission cumulative plots of 15\% and 85\% are shown in the
lower panel of Fig.~\ref{fig_fixed_structure}. The cumulative plots
show that the continuum emission up to $\sim$100 $\mu$m comes from
within the first 20~AU of the disc. Most of the \OIfs\ emission is
confined within 20~AU as well.

For the vertical hydrostatic discs, none of the models, either compact
($R_{\mathrm {out}}$=15~AU) or extended ($R_{\mathrm {out}}$=400~AU),
managed to simultaneously fit the SED from near-IR to the millimeter
flux density. The best simultaneous fit to the SED from the near- to
the far-IR, excluding the data point at 1.2~mm and \OIfs\ flux, is
obtained for a compact disc (model 1 and 2) with a disc gas mass of
5$\times$10$^{-6}$ M$_\odot$ and a dust mass of 5$\times$10$^{-8}$
M$_\odot$. The fits to the SED with and without dust settling do not
fit the observed continuum emission at 1.2mm. The models underpredict
the 1.2~mm flux because of the small dust mass in the models.  Models
7 and 8, both with dust mass of 10$^{-6}$ M$_\odot$ as suggested by
the simple analysis of the 1.2~mm flux, overpredict all observed
fluxes at wavelengths shorter than 1.2~mm. The gas scale-heights in
the hydrostatic models are much higher than the value suggested by the
best model with a fixed vertical structure disc. As a result, even
with dust settling, the hydrostatic models of 400~AU discs produce far
too strong far-IR continuum fluxes.

In the absence of spatially resolved image at 1.2~mm, we can constrain
neither the location of the cold dust grains nor their geometry
precisely. However, an extended flat disc has been advocated by
\citet{Stark2009ApJ...703.1188S}, who fitted the SED and mid-IR
interferometry data simultaneously.  An alternative explanation for
the 1.2~mm emission is stellar free-free emission, which can also
explain the H$\alpha$ and Br$\gamma$ emission.

Hydrostatic disc models with dust settling provide a better fit of the
mid- and far-IR photometric points while the model without dust
settling performs better in the near-IR but less well in the
far-IR. This conclusion on the settling applies only for the
assumption that the grains are composed of pure compact spherical
amorphous silicate grains. Other grain compositions and shapes are
possible. Grains with amorphous carbon inclusions emit more
efficiently in the near-infrared. The \OIfs\ flux from the
unsettle-model is closer to the observed value but both flux estimates
are well within the observational uncertainties. We tested the effect
of well-mixed gas and dust (no settling) model and all other
parameters the same. The fit to the SED of the well-mixed model is
slightly worse (overpredicting the flux in the far-IR) than in the
model with settling and the [OI] emissions are comparable, albeit
different (Table~\ref{model_fluxes}).

The disc vertical hydrostatic structure shown in
Fig.~\ref{fig_hydrostatic_density} is complex but clearly drops beyond
8~AU due a drop in the gas temperature in both the settled and
well-mixed models. The dust density structure follows that of the gas
structure (Fig.~\ref{fig_hydrostatic_dust_structure}). The dust grains
are warmer farther out in the settled-dust model, which is expected
since the grains are on average smaller in the disc upper layers.  We
also show the gas temperature in Fig.~\ref{fig_hydrostatic_Tgas},
which is much higher than the dust temperature in the disc atmosphere.

The [OI] flux from models 1 and 2 slightly unpredicts the observed
flux. The [OI] flux emission area is shown in
Fig.~\ref{fig_hydrostatic_OI63_cumulative}. The \OIfs\ emission at 63
micron arises from the disc midplane within the first 10-15~AU. The
\OIfs\ line does probably not probe the same circumstellar region as
the 1.2~mm continuum emission. It is clear that the \OIfs\ line cannot
be used to probe cold gas emission.

In all models, atomic oxygen is by far the main oxygen carrier in the
disc, followed by CO. The disc midplane is molecular and CO reaches
its maximum abundance of 1.3$\times$10$^{-4}$. In our models the
\OIfs\ emission traces the disc molecular gas. The disc around
\object{51 Oph} is mostly molecular (H$_2$, CO, etc) and poor in C$^+$
apart from the upper disc atmospheres. The resulting \CII\ fluxes are
low, consistent with the observed upper limit. There is an
intermediate layer where neutral atomic carbon is abundant. The oxygen
63 micron line is emitted within 15~AU of the star regardless of the
physical size of the disc.
  
The main gas coolant in the \OIfs\ emitting region are the CO
ro-vibrational lines.  The \OIfs\ line probes a tiny fraction of the
total disc gas mass only. In the absence of PAHs the main heating is
the CO IR-pumping and background heating by atoms and ions (FeII ,
SiII, CII, CI) in the first 5 AU and dust photoelectric on dust grains
beyond 5 AU. Water is a very minor component, and its ro-vibrational
lines do not contribute to the gas cooling.

A larger disc ($R_{\mathrm{out}}$=400 AU) with the same gas mass as
the best compact model predicts the continuum and \OIfs\ line
flux. The modelling confirms that we cannot constrain the actual disc
outer radius. However, the location of the \OI\ emission is confined
to the first 10-15~AU for both the compact and the more extended
disc. The more massive large discs (models 7 and 8) do not match the
SED and overpredict the \OIfs\ flux. A few models with sub-standard
gas-to-dust mass ratio have \OIfs\ flux close to the observed
values. However, like all other hydrostatic model, the fits to the SED
are poor.

\section{Discussion}\label{discussion}

What is the origin of the gas and dust grains in the inner disc?  Do
the grains come from the coagulation of interstellar grains, or are
they the results of collisions between planetesimals scattered from
the outer belt reservoir by unseen companions? Is the gas the remnant
of an initial massive disc? \object{51 Oph} is young: $\sim$0.7~Myr
\citep{Montesinos2009A&A...495..901M}, while the youngest stars with
established debris discs are about 10 Myr old (\object{$\beta$ Pic}
has an age of ~12 Myr). Could the disc around \object{51 Oph} be
considered as a debris disc at its age?

The best fixed vertical structure model is a flat gas disc whereas the
gas scale height derived from the hydrostatic models is much more
extended vertically. The comparison between the fixed vertical
structure disc model and the hydrostatic models suggests that there
may be a mechanism to flatten the gas disc around \object{51~Oph}.
The vertical hydrostatic structure is derived from balancing the
thermal pressure and the gravity forces.

  The disc average gas-to-dust mass ratio is $\sim$~30--100 for the
  best model, consistent with the standard interstellar value and with
  the ratio in other Herbig Ae discs: \object{HD~169142}
  \citep{Meeus2010A&A...518L.124M} and \object{HD~163296}
  \citep{Tilling2012A&A...538A..20T}. A gas-to-dust mass ratio of 100
  is much higher than the ratio in the debris discs
  \citep[e.g.,][]{Lebreton2012A&A...539A..17L}, where the dust grains
  come from shattering of planetesimals and the gas is
  absent. Nevertheless, the \object{51~Oph} disc gas mass
  (5$\times$10$^{-6}$ M$_\odot$) is much lower than the gas mass of
  the aforementioned HerbigAe discs, which can reach 0.01
  M$_\odot$. If the disc around \object{51 Oph} is primordial, as
  suggested by its young age, then either the initial disc had a low
  mass or the initial massive disc has been actively dissipating
  either by photoevaporation, accretion onto the star, and/or planet
  formation. As discussed above, the current mass accretion rate onto
  \object{51 Oph} is difficult to assess, not to mention the accretion
  rate history. In discs where the gas is replenished by planetesimal
  evaporations, the gas-to-dust ratio would be closer to
  unity. Hydrodynamic simulations show that dust grains can gather
  into eccentric rings as observed in near-IR images of debris discs
  without the need of the presence of a planet
  \citep{Lyra2012arXiv1204.6322L}.

  The high luminosity of the star ($L_*$=~260-300 $L_\odot$) compared
  with other HerbigAe stars ($L_*$=~20-40 $L_\odot$) may explain a
  fast disc photoevaporation or increases in the efficiency of
  radiative pressure on grains resulting in a radially extended
  disc. Disc photo-evaporation occurs when hot gases have enough
  energy to escape the star+disc gravitational potential entraining
  the dust grains that are dynamically coupled to the gas. The upper
  disc atmosphere outside the critical radius of $\sim$6.8 AU for
  \object{51 Oph} is unbound and flows as a wind
  \citep{Hollenbach1994ApJ...428..654H}. The mass-loss rate through
  disc evaporative wind is $\dot{M}_{\mathrm{w}}\simeq
  4\times10^{-10}$ M$_\odot$ yr$^{-1}$ assuming a stellar ionizing
  flux of $\Phi$ = 10$^{42}$ ionizing photons per second, valid for a
  low-mass solar-luminosity star
  \citep{Alexander2009ApJ...704..989A,Pascucci2011ApJ...736...13P}. Scaling
  to the luminosity of \object{51~Oph}, the wind mass-loss rate
  reaches 1.2$\times$10$^{-7}$ yr$^{-1}$. A detailed disc-wind model
  is needed to test whether the mass-loss rate of $\sim$10$^{-7}$
  yr$^{-1}$ can explain the observed H$\alpha$ and Br$\gamma$ line
  fluxes and profiles. In 0.7~Myr, the disc around \object{51~Oph}
  would have lost 0.1 M$_\odot$ of gas. If photoevaporation is
  occurring at a constant rate, the initial disc gas mass around
  \object{51~Oph} would have been $\sim$0.1 M$_{\odot}$ compared with
  the current stellar mass of 2.8 M$_\odot$. Since the inner disc is
  warm, the initial disc would have been gravitationally stable. The
  current disc within 10-15~AU will quickly disappear at this high
  photoevaporation rate. A mass-loss rate of $\sim$10$^{-7}$ yr$^{-1}$
  is very high compared with the mass in in the disc
  ($\sim$~5$\times$10$^{-6}$ M$_\odot$) and may be overestimated. A
  detailed disc-wind model is warranted to better constrain the
  photoevaporative mass-loss rate.

  The dust mass in the small compact disc around \object{51~Oph}
  derived from the simultaneous fit to the SED (near- to far-IR
  photometry data) using the hydrostatic disc model is small at
  5$\times$10$^{-8}$ M$_\odot$ (0.017 M$_\oplus$), two orders of
  magnitude lower than if all the 1.2~mm flux were attributed to dust
  emission. However, this dust mass is still ten times the value
  derived by \citet{vandenAncker2001A&A...369L..17V} Their value was
  derived from a fit to photometric points at wavelengths shorter than
  100 microns only.  Our derived dust mass is five times higher than
  the mass found by \citet{Tatulli2008A&A...489.1151T}.

  The discrepancies can also be attributed to differences in dust
  opacities. The continuum emission at 63 micron is optically thin,
  thus it probes the amount of cool dust in the disc.

  The best models with a fixed vertical structure or with a disc in
  vertical hydrostatic equilibrium suggest that the gas-to-dust mass
  ratio is $\sim$~30--100, but are the gas and dust well-mixed or have
  the dust grains settled to the midplane? We used the model results
  to compute the Stokes parameter
  \citep{Weidenschilling1977MNRAS.180...57W},
\begin{equation}
  S_t = \frac{\Omega_K}{2\pi}\frac{a_d m_d}{c_s\rho_g},
\end{equation}
where $\Omega_K=(GM_*/R^3)^{1/2}$ is the Keplerian angular velocity at
radius $R$, $a_d$ the grain radius in cm, $m_d$ is the volume mass
density of a dust grain, which is $\sim$1 g cm$^{-3}$ for a silicate
grain, $c_s$ the sound speed in cm s$^{-1}$, and
$\rho_g=n_{\mathrm{H}}\times$ a.m.u. the gas mass density in g
cm$^{-3}$ (a.m.u. is the atomic mass unit in grams). The Stokes
parameter varies throughout the disc. It measures the level of dust
grain coupling with the gas via drag force. For $S_t\gg$1, the gas
drag is inefficient and the dust grains quickly settle towards the
disc midplane. For $S_t\ll$1, the gas and dust grains are well mixed
and are collocated. The contours for $\log_{10}$ the Stokes parameters
(values of -10, 1, and 10) in the disc models with and without
settling for the mean grain size are plotted over the density
distribution in Fig.~\ref{fig_hydrostatic_density} and over the dust
density distribution in Fig.~\ref{fig_hydrostatic_dust_structure}. The
Stokes parameter is small throughout the disc in our models because
the gas is dense ($n_{\mathrm{H}}>10^{6}$ cm$^{-3}$) and warm
($T_{\mathrm{gas}}>$~250~K).
  
Fits to the SED and gas lines do not strongly constrain the presence
of dust settling.  However, the SED together with VLT-MIDI and Keck
nulling data support models with dust settling
\citep{Stark2009ApJ...703.1188S}. A more detailed modelling is needed
to compute the exact dust settling, coagulation, and migration, which
is beyond the scope of this paper
\citep[e.g.][]{Fouchet2005A&A...443..185B}.

Interestingly, dust settling does not much affect the gas vertical
hydrostatic structure for two reasons First, photoelectric-heating by
dust grains is not the only heating process in the disc layers located
above the midplane. Heating by PAHs, which are small enough to be
coupled to the gas, is important, but their abundance is very
low. Second, sub-micron grains, which are more strongly coupled to the
gas and remain in the disc atmosphere, provide most of the
photoelectrons. Likewise, the gas line collisional excitations do not
change much in our model of the \object{51~Oph} disc whether the dust
grains have settled or not.

If we can attribute the millimeter emission to a cold outer disc
instead of free-free emission and the near- to mid-IR continuum
emission and the \OIfs\ to a warm compact inner disc, the
circumstellar environment of \object{51 Oph} resembles a young and
massive counterpart to main-sequence stars with a warm exozodiacal
dust ring and a cold outer planetesimal belt
\citep{Defrere2012A&A...546L...9D}.  The main difference between
\object{51~Oph} and \object{$\beta$ Pic} is that the dust grains in an
exozodiacal ring around \object{$\beta$ Pic} have to be replenished,
while part the dust grains around \object{51 Oph} may still be
primordial.

\section{Conclusions}\label{conclusions}

We detected \OIfs\ far-IR and 1.2~mm continuum emissions towards
\object{51~Oph}. The \OIfs\ probes the warm gas in a compact inner
circumstellar disc while the 1.2~mm emission suggests a cold outer
disc. An extended flat disc with dust settling can explain the
continuum and line data consistent with previous model. Hydrostatic
disc models predict discs that are too extended in the vertical
direction for the gas and the dust. The failure of the hydrostatic
models to explain the observations suggests that mechanism(s) other
than thermal pressure and gravity forces are at work to keep the gas
and dust disc flat. The gas-to-dust mass ratio is consistent with the
standard value of 100. Future observations in the (sub)-millimeter
with ALMA will help constraining the cold gas and dust around
\object{51 Oph}.

\begin{acknowledgements}
  We thank the Agence Nationale pour la Recherche (contracts
  ANR-07-BLAN-0221, ANR-2010-JCJC-0504-01, and ANR-2010-JCJC-0501-01),
  CNES, PNPS of CNRS/INSU, France, and support from the Millenium
  Science Initiative (Chilean Ministry of Economy), through grant
  ''Nucleus P10-022-F'' for support. We acknowledge funding from the
  EU FP7-2011 under Grant Agreement nr. 284405
  (PERG06-GA-2009-256513).  Computations presented in this paper were
  performed at the Service Commun de Calcul Intensif de l'Observatoire
  de Grenoble (SCCI) on the super-computer Fostino funded by
  ANR. C. Eiroa, G. Meeus, and B. Montesinos are partly supported by
  Spanish grant AYA 2011-26202. We finally thank the referee for
  his/her comments that helped improving the manuscript.
\end{acknowledgements}

\bibliographystyle{aa} 
\bibliography{herschel51Oph.bib}

\begin{appendix}

\section{Collisional data}\label{collisional_rates}

The original articles for the line frequencies, Einstein coefficients,
and collisional rates are CO
\citep{Flower2001JPhB...34.2731F,Jankowski2005JChPh.123j4301J,Yang2006JChPh.124j4304Y,Wernli2006A&A...446..367W},
H$_2$O
\citep{Barber2006MNRAS.368.1087B,Dubernet2002A&A...390..793D,Faure2004MNRAS.347..323F,Faure2007A&A...472.1029F,Daniel2011A&A...536A..76D},
\OI\
\citep{Abrahamsson2007ApJ...654.1171A,Bell1998MNRAS.293L..83B,Chambaud1980JPhB...13.4205C,Jaquet1992JPhB...25..285J,Launay1977JPhB...10..879L},
\CII\
\citep{Flower1977JPhB...10.3673F,Launay1977JPhB...10..879L,Wilson2002MNRAS.337.1027W},
CH$^+$
\citep{Muller2010A&A...514L...6M,Lim1999MNRAS.306..473L,Hammami2009A&A...507.1083H,turpin2010A&A...511A..28T},
OH \citep{Offer1994JChPh.100..362O}.

\end{appendix}

\end{document}